\documentclass[aps,
preprint,
eqsecnum,
tightenlines,floats,
]{revtex4-1}
\usepackage{amssymb}
\usepackage{amsmath}
\usepackage{bm}
\usepackage{hyperref}
\usepackage{graphicx}
\usepackage{subfigure}

\begin{document}

\date{\today}

\title{Density perturbations in axion-like particles: classical vs quantum field treatment}
\author{Sankha Subhra Chakrabarty$^{a,b}$}
\affiliation{
$^a$Dipartimento di Fisica, Universit\`a di Torino, Via P. Giuria 1, I-10125 Torino, Italy \\
$^b$Istituto Nazionale di Fisica Nucleare (INFN), Sezione di Torino, Via P. Giuria 1, I-10125 Torino, Italy}

\begin{abstract}

Axions and axion-like particles are bosonic quantum fields. They are often assumed to follow classical field equations due to their high degeneracy in the phase space. In this work, we explore the disparity between classical and quantum field treatments in the context of density and velocity fields of axions. Once the initial density and velocity field are specified, the evolution of the axion fluid is unique in the classical field treatment. However, in the quantum field treatment, there are many quantum states consistent with the given initial density and velocity field. We show that evolutions of the density perturbations for these quantum states are not necessarily identical and, in general, differ from the unique classical evolution. To illustrate the underlying physics, we consider a system of large number of bosons in a one-dimensional box, moving under the gravitational potential of a heavy static point-mass. We ignore the self-interactions between the bosons here. Starting with homogeneous number density and zero velocity field, we determine the density perturbations in the linear regime in both quantum and classical field theories. We find that classical and quantum evolutions are identical in the linear regime if only one single-particle state is occupied by all the bosons and the self-interaction is absent. If more than one single-particle states are occupied, the density perturbations in quantum evolutions differ from the classical prediction after a certain time which depends upon the parameters of the system. 

\end{abstract}

\maketitle

\section{\label{I}Introduction}

Axions or axion-like particles are highly motivated candidates for dark matter (see e.g. \cite{Preskill:1982cy, Abbott:1982af, Dine:1982ah, Hu:2000ke, Hui:2016ltb, deMartino:2020gfi}). Although they are bosonic quantum fields, in the context of cosmology and large scale structure formation, they are often approximated as classical scalar fields due to their high degeneracy in the phase space \cite{Sin:1992bg,  Hu:2000ke, Mielke:2009zza, Lora:2011yc, Marsh:2013ywa, Schive:2014dra, Li:2013nal, Hui:2016ltb, Lee:2008jp, Lundgren:2010sp,Marsh:2010wq,RindlerDaller:2011kx,Chavanis:2016dab, Vaquero:2018tib, Veltmaat:2018dfz, Hui:2020hbq, Li:2018kyk}. In the non-relativistic regime, the classical field approximation leads to a wave description of the axions where the complex wavefunction satisfies the Schr\"odinger-Gross-Pitaevskii (SGP) equation. The wave description can be mapped into a classical fluid description of axions where number density and velocity field of the axions follow the continuity equation and the Euler equation with an additional `quantum pressure' term (see e.g., \cite{Chakrabarty:2017fkd, Li:2018kyk}).

It has been argued that the classical field approximation may not provide an accurate description of axions with contact or gravitational self-interactions, especially in the context of thermalization of axions \cite{Sikivie:2009qn, Erken:2011dz, Saikawa:2012uk, Sikivie:2016enz, Chakrabarty:2017fkd, Arza:2019kab}. There is a time-scale after which the classical results significantly differ from the results of quantum field treatment. Since thermalization of axions was the central issue, most of the previous studies focussed on how the occupation numbers evolve in quantum and classical field theories. In this work, we investigate whether the density and velocity field of axions evolve differently in quantum and classical field treatments. 

In the classical field approximation, the evolution of axion fluid is uniquely specified by initial values of the number density $n(\vec x,0)$ and the velocity field $\vec v (\vec x, 0)$. In the quantum field treatment, the evolution of the system is uniquely specified only by the initial state of the system. We show that the evolution of axion number density in the quantum field description is not unique for given initial values of number density $n(\vec x,0)$ and velocity field $\vec v (\vec x, 0)$. There exist a large number of quantum states corresponding to the given values of $n(\vec x,0)$ and $\vec v (\vec x, 0)$. Evolutions of number densities for all these quantum states are not identical and, in general, differ from the unique classical evolution. We argue that, for certain quantum states, the classical and quantum evolutions of density may be different, even in the absence of self-interactions among axions.

In Sec.~\ref{II}, we provide a formalism to determine the evolution of number density in the quantum field treatment. We explain why quantum and classical field treatments of a system of axions may yield different results within the linear regime. In Sec.~\ref{III}, we consider a toy-model of axions in a one-dimensional box. Using both classical and quantum field theories, we find the growth of density perturbations under the gravitational potential of a static point mass. We show that, depending upon the parameters and the quantum state of the system, the quantum field treatment may result into significantly different outcome as compared to the classical result. Section~\ref{IV} provides a summary.

\section{\label{II}Evolution of density and velocity}

Axions and axion-like particles are quantum scalar field with mass $m$. In the non-relativistic regime, they are described by a complex scalar field $\psi (\vec x , t)$ satisfying the Schr\"odinger-Gross-Pitaevskii (SGP) equation (see e.g. \cite{Chakrabarty:2017fkd}):
\begin{equation}
i \partial _t \psi (\vec x,t) = \tfrac{(-1)}{2m} \ \nabla ^2 \psi (\vec x,t) + m \hat V[\psi (\vec x,t)] \ \psi (\vec x,t) \label{eq:SGP_q}
\end{equation}
where  $\hat V$ is the potential operator. In general, it is a functional of the field $\psi(\vec x,t)$ which satisfies the equal-time commutation relations: $[\psi (\vec x,t) , \psi (\vec y,t)] = 0$ and $[\psi (\vec x,t) , \psi^\dagger (\vec y,t)] = \delta^{(3)}(\vec x - \vec y)$. In quantum field theory, number density $n(\vec x,t)$ and the current density $\vec j (\vec x,t)$ are given by the expectation values of the following operators:
\begin{eqnarray}
\hat n_Q (\vec x,t) &=& \psi^\dagger (\vec x,t) \psi(\vec x,t) \nonumber \\
\hat{\vec j}_Q (\vec x,t) &=& \tfrac{1}{2im} \left( \psi^\dagger (\vec x,t) \vec \nabla \psi (\vec x,t)  - \vec \nabla \psi^\dagger (\vec x,t) \psi (\vec x,t) \right) \ . \label{eq:nj_q}
\end{eqnarray}
The velocity field $\vec v_Q (\vec x,t)$ is given by $\frac{\langle \hat{\vec j}_Q (\vec x,t) \rangle}{\langle \hat n_Q(\vec x,t) \rangle}$. Since we are working in the Heisenberg picture, the quantum state of the system does not evolve. For example, if the initial quantum state is $\left| {\rm A} \right>$, the number density at time $t$ is given by $\left< {\rm A} \right| \hat n_Q (\vec x,t) \left| {\rm A} \right>$. Using Eq.~\ref{eq:SGP_q} and \ref{eq:nj_q}, we get
\begin{eqnarray}
\partial_t \left < \hat n_Q \right> + \vec \nabla \cdot \langle \hat{\vec j}_Q \rangle &=& 0  \label{eq:cont_q} \\
\partial_t \langle \hat{\vec j}_Q \rangle + \langle \psi^\dagger (\vec \nabla \hat V) \psi \rangle &=& \langle\hat{\vec F}_Q \rangle  \label{eq:Euler_q}
\end{eqnarray}
where
\begin{eqnarray}
\hat{\vec F}_Q = \tfrac{1}{4m^2} [&&\psi^\dagger(\vec \nabla \nabla^2 \psi) + (\vec \nabla \nabla^2 \psi^\dagger) \psi \nonumber \\
&&- (\nabla^2 \psi^\dagger)(\vec \nabla \psi) - (\vec \nabla \psi^\dagger)(\nabla^2 \psi)] \ .
\end{eqnarray}
In general, $\langle \psi^\dagger (\vec \nabla \hat V) \psi \rangle \neq \langle \hat n_Q \rangle \langle \vec \nabla \hat V \rangle$ and $\langle \hat{\vec F}_Q \rangle$ cannot be expressed in terms of $\langle \hat n_Q \rangle$ and $\langle \hat{\vec j}_Q\rangle$. Therefore, the initial values of $\langle \hat n_Q \rangle$ and $\langle \hat{\vec j}_Q\rangle$ cannot determine their evolutions. Evolutions of $\langle \hat n_Q \rangle$ and $\langle \hat{\vec j}_Q\rangle$ are uniquely determined only if the initial quantum state of the system is known, as we show in Sec.~\ref{dens_Q}.

In the commonly used classical field treatment, $\psi(\vec x,t)$ in Eq.~\ref{eq:SGP_q} is replaced by a complex wavefunction $\Psi (\vec x,t)$ which satisfies the SGP euqtion:
\begin{equation}
i \partial _t \Psi (\vec x,t) = \frac{(-1)}{2m} \ \nabla ^2 \Psi (\vec x,t) + m V[\Psi (\vec x,t)] \ \Psi (\vec x,t) \ . \label{eq:SGP_cl}
\end{equation}
Writing $\Psi (\vec x,t) = |\Psi (\vec x,t)| e^{i\beta(\vec x,t)}$, one can map the axions into a fluid with number density $n_{\rm cl} (\vec x,t) = |\Psi (\vec x,t)|^2$ and velocity field $\vec v_{\rm cl} (\vec x,t) = \frac{1}{m} \vec \nabla \beta (\vec x,t)$. The classical SGP equation implies that $n_{\rm cl} (\vec x,t)$ and $\vec v_{\rm cl} (\vec x,t)$ satisfy the continuity equation and the Euler equation with an additional `quantum pressure' term (see e.g. Ref.~\cite{Chakrabarty:2017fkd}):
\begin{eqnarray}
\partial_t n_{\rm cl} + \vec \nabla \cdot (n_{\rm cl} \vec v_{\rm cl}) &=& 0 \nonumber \\
\partial_t \vec v_{\rm cl} + (\vec v_{\rm cl} \cdot \vec \nabla) \vec v_{\rm cl} &=& - \vec \nabla V - \vec \nabla q \label{eq:class}
\end{eqnarray}
where 
\begin{equation}
q = \frac{(-1)}{2m^2} \frac{\nabla^2 \sqrt{n_{\rm cl}}}{\sqrt{n_{\rm cl}}} \ . \label{eq:qclass}
\end{equation}
Since the equations are first order in time, the number density $n_{\rm cl} (\vec x,t)$ and the velocity field $\vec v_{\rm cl} (\vec x,t)$ are uniquely determined by the initial values, $n_{\rm cl} (\vec x,0)$ and $\vec v_{\rm cl} (\vec x,0)$.

In the quantum theory, for given $n(\vec x,0)$ and $ \vec j (\vec x,0)$, there exist a set of initial quantum states, $S \equiv \{ \left| {\rm in} \right>\}$ such that each state in $S$ satisfies
\begin{eqnarray}
\left< {\rm in} \right| \hat n_Q (\vec x,0) \left| {\rm in} \right> &=& n(\vec x,0) \nonumber \\
\left< {\rm in} \right| \hat{\vec j}_Q (\vec x,0) \left| {\rm in} \right> &=& \vec j (\vec x,0) \ .
\end{eqnarray}
We explore whether the evolutions of number densities corresponding to all the initial quantum states in $S$ are identical. In Sec.~\ref{dens}, we explain how the number density is related to the quantum state of the system of bosons. In Sec.~\ref{dens_Q} and \ref{dens_C}, we describe how to find the evolution of the density in quantum and classical field theories, respectively, by treating the potential as a perturbation.

\subsection{\label{dens}Number density in quantum field treatment}

The most fundamental difference between classical and quantum field treatments is that $\Psi (\vec x,t)$ in the classical field theory is a complex wavefunction, whereas $\psi (\vec x,t)$ in the quantum field theory is an operator which acts on the state of the system of axions. The number density is given by $|\Psi (\vec x,t)|^2$ in the classical field theory. In quantum field theory, it is given by the expectation value of $\hat n (\vec x,t) = \psi^\dagger (\vec x,t) \psi (\vec x,t)$. 

The quantum field $\psi(\vec x,t)$ is expanded in terms of a set of orthonormal and complete functions $u^{\vec k} (\vec x,t)$:
\begin{equation}
\psi(\vec x,t) = \sum_{\vec k} a_{\vec k} (t) u^{\vec k} (\vec x,t) \label{eq:psi}
\end{equation}
where the annihilation operators $a_{\vec k} (t)$ satisfy $[a_{\vec k} (t) , a_{\vec k'} ^\dagger (t)] = \delta^{\vec k'} _{\vec k}$, and $u^{\vec k} (\vec x,t)$'s satisfy:
\begin{eqnarray}
\sum_{\vec k} u^{\vec k} (\vec x,t) u^{\vec k} (\vec y,t)^* = \delta^{(3)}(\vec x - \vec y) \nonumber \\
\int d^3 x \; u^{\vec k} (\vec x,t)^* u^{\vec k'} (\vec x,t) =  \delta^{\vec k'} _{\vec k} \; \; . \label{eq:uk_cond}
\end{eqnarray}
$u^{\vec k} (\vec x, t)$ can be any set of complete orthonormal functions e.g., the momentum eigenstates, the eigenstates of an Hamiltonian. In the rest of the paper, we assume $u^{\vec k}$'s to be time-independent.

The quantum state of a system of $N$ identical bosons can be written as
\begin{equation}
\left| \rm in \right> = \sum_{\{j\}} c_{\{j\}} \left| \{ N_1 , N_2, ... , N_M \}_j \right> \label{eq:qstate}
\end{equation}
where the summation is over all possible combinations of placing N identical particles in $M$ single-particle states: $\{u^{\vec k_1} (\vec x) , u^{\vec k_2} (\vec x) , ... , u^{\vec k_M} (\vec x) \}$  and $(N_1 + N_2 + ... + N_M) = N$. Let us consider one such combination, $\left| N_1 , N_2, ... , N_M \right>$, where $N_1$ particles are in the single-particle state $u^{\vec k_1} (\vec x)$, $N_2$ particles in $u^{\vec k_2} (\vec x)$ and so on. If this is the quantum state at time $t = t_0$, then
\begin{equation}
\left| N_1 , ... , N_M \right> = \left(\prod_{i=1}^{M}\frac{1}{\sqrt{N_i !} } \ (a^\dagger _{\vec k_i} (t_0))^{N_i} \right) \left| 0 \right>  
\end{equation}
where $\left| 0 \right> $ represents the vacuum state. It is straightforward to show that
\begin{eqnarray}
a_{\vec k} (t_0) \left| N_1 , ... , N_M \right> &=& \delta_{\vec k} ^{\vec k_1} \sqrt{N_1} \left| N_1 - 1 , ... , N_M \right> \nonumber \\
&&+ ... + \delta_{\vec k} ^{\vec k_M} \sqrt{N_M} \left| N_1 , ... , N_M -1 \right> 
\end{eqnarray}
and, the operation of $\psi (\vec x,t_0)$ yields
\begin{eqnarray}
\psi (\vec x,t_0)  \left| N_1 , ... , N_M \right> &=& \sqrt{N_1} \ u^{\vec k_1} (\vec x) \left| N_1 -1 , ... , N_M \right> \nonumber \\
&&+ ... + \sqrt{N_M} \ u^{\vec k_M} (\vec x) \left| N_1 , ... , N_M -1 \right> \ \ . 
\end{eqnarray}
Therefore, the number density corresponding to the state $\left| N_1 , ... , N_M \right>$ at time $t_0$ is
\begin{eqnarray}
n_Q (\vec x, t_0) &=& \left < N_1, ..., N_M \right | \psi^\dagger (\vec x,t_0) \psi (\vec x,t_0) \left | N_1, ..., N_M \right> \nonumber \\
&=& N_1 \left | u^{\vec k_1} (\vec x) \right |^2 + ... + N_M \left | u^{\vec k_M} (\vec x) \right |^2 \ . \label{eq:nQt0}
\end{eqnarray}
We emphasize two following points: the above expression is the number density corresponding to an eigenstate of occupation number operator and $u^{\vec k} (\vec x)$'s can be any set of complete orthonormal functions satisfying Eqs.~\ref{eq:uk_cond}. Now we discuss how homogeneous and inhomogeneous number densities are represented in the framework of quantum field theory. We also point out an inconsistency between quantum field theory and the commonly used classical field treatment when more than one single-particle states are occupied.

\subsubsection{\label{homogeneous} Homogeneous number density}

If we choose $u^{\vec k}$'s to be the momentum eigenstates: $u^{\vec k} (\vec x) = \tfrac{1}{\sqrt{V}} \ e^{-i \vec k \cdot \vec x} $, the number density corresponding to the state $\left| A \right> = \left| N_1 , ... , N_M \right>$ turns out to be
\begin{equation}
\left < N_1, ..., N_M \right | \psi^\dagger (\vec x,t_0) \psi (\vec x,t_0) \left | N_1, ..., N_M \right> = \frac{N}{V}
\end{equation}
following Eq.~\ref{eq:nQt0}. A homogeneous fluid consisting of $N_i$ particles in momentum eigenstates $\vec k_i$ ($i = 1,2, ... , M$) may appear to be counter-intuitive. However, from the viewpoint of quantum mechanics, each particle is in one of the momentum eigenstates and, as a result, has maximum uncertainty in position which leads to the homogeneity.

One can also arrive at this conclusion by considering a many-particle wavefunction of the bosons. We are explaining in terms of the simplest scenario with two bosons where one is in momentum eigenstate $u^{\vec k_1}(\vec x) = \tfrac{1}{\sqrt{V}} \ e^{-i \vec k_1 \cdot \vec x}$ and the other at $u^{\vec k_2} (\vec x) = \tfrac{1}{\sqrt{V}} \ e^{-i \vec k_2 \cdot \vec x}$. The symmetrized wavefunction is 
\begin{equation}
\Psi (\vec x_1, \vec x_2) = \frac{1}{\sqrt{2!}}\left(u^{\vec k_1}(\vec x_1) u^{\vec k_2}(\vec x_2) + u^{\vec k_1}(\vec x_2) u^{\vec k_2}(\vec x_1)   \right)
\end{equation}
The number density at $\vec x$ is given by 
\begin{equation}
n(\vec x) = \sum_{j=1}^{2} \int d^3 x_1 \ d^3 x_2 \ \delta^{(3)} (\vec x - \vec x_j) |\Psi (\vec x_1, \vec x_2)|^2 = \frac{2}{V}
\end{equation}
when $\vec k_1 \neq \vec k_2$. Generalization of this into larger number of bosons is possible but tedious due to the symmetrization. In fact, this is where a quantum field framework is advantageous.

\subsubsection{\label{inhomogeneous} Inhomogeneous number density}

To explain how the  quantum state of an inhomogeneous fluid may be written in terms of momentum eigenstates, let us consider the following quantum state:
\begin{equation}
\left| B \right> = c_1 \left| N_1 , N_2 \right> + c_2 \left| N_1 -1 , N_2 + 1 \right>
\end{equation}
where $c_1$ and $c_2$ are complex coefficients satisfying $|c_1|^2 + |c_2|^2 = 1$ and $(N_1 + N_2)=N$. Number of particles in states other than $u^{\vec k_1} (\vec x)$ and $u^{\vec k_2} (\vec x)$ are zero. The result of $\psi (\vec x,t_0)$ acting on $\left| B \right>$ is
\begin{eqnarray}
\psi (\vec x,t_0) \left| B \right> &&= \tfrac{1}{\sqrt{V}} \big[c_1 e^{-i \vec k_1 \cdot \vec x} \sqrt{N_1} \left| N_1 -1 , N_2 \right> \nonumber \\
&&+ c_1 e^{-i \vec k_2 \cdot \vec x} \sqrt{N_2} \left| N_1 , N_2 -1 \right> \nonumber \\
&&+ c_2 e^{-i \vec k_1 \cdot \vec x} \sqrt{N_1 - 1} \left| N_1 -2 , N_2 + 1 \right> \nonumber \\
&&+ c_2 e^{-i \vec k_2 \cdot \vec x} \sqrt{N_2 + 1} \left| N_1 -1 , N_2 \right> \big] \ . \label{eq:psi_B}
\end{eqnarray}
The first and the last terms of Eq.~\ref{eq:psi_B} interfere and result into a mode of density perturbation with wavevector $\vec k_{21} = (\vec k_2 - \vec k_1)$:
\begin{eqnarray}
&&n_Q (\vec x,t_0) = \left< B \right| \psi^\dagger (\vec x,t_0) \psi (\vec x,t_0) \left| B \right> \nonumber \\
&&= \tfrac{N}{V} \left[1 + 2 |c_1| |c_2| \sqrt{\tfrac{N_1}{N} \tfrac{(N_2 + 1)}{N}} \cos (\vec k_{21} \cdot \vec x + \delta_{21}) \right] 
\end{eqnarray}
where $\delta_{21} = \delta_2 - \delta_1$ and $c_l = |c_l| e^{-i \delta_l}$ ($l = 1,2$). In general, density perturbation in mode $\vec k$ is caused by the exchange of a boson between a pair of momentum-eigenstates $u^{\vec k_i}$ and $u^{\vec k_j}$, provided $(\vec k_j - \vec k_i ) = \vec k $ or $(-\vec k)$.

\subsubsection{\label{inconsistency} An inconsistency between quantum field treatment and the commonly used classical theory}

In the commonly used classical field treatment, the wavefunction is expressed as $\Psi (\vec x, t) = \frac{1}{\sqrt{V}} \sum_{\vec k} A_{\vec k} (t) e^{-i\vec k \cdot \vec x}$. Then the occupation number in the eigenstate of momentum $\vec k$ is defined as $N_{\vec k} (t) = A_{\vec k}^* (t) A_{\vec k} (t)$. In this spirit, one would express the classical wavefunction corresponding to the quantum state $\left|N_1(\vec k_1), N_2 (\vec k_2) \right>$ as
\begin{equation}
\Psi (\vec x) = \frac{1}{\sqrt{V}} \left[ \sqrt{N_1} e^{-i\vec k_1 \cdot \vec x} + \sqrt{N_2} e^{-i\vec k_2 \cdot \vec x} \right] \ .
\end{equation}
The corresponding number density in the classical field theory is $n_{\rm cl} (\vec x) = |\Psi (\vec x)|^2$ which is inhomogeneous. However, the number density in the quantum field theory corresponding to the quantum state $\left|N_1(\vec k_1), N_2 (\vec k_2) \right>$ is homogeneous as shown in Sec.~\ref{homogeneous}.

The reason behind this apparent inconsistency is that $N_{\vec k} (t) = A_{\vec k}^* (t) A_{\vec k} (t)$ does not represent the number of particles in state $\vec k$. This should come as no surprise because the SGP equation describes the evolution of the single-particle state $\Psi (\vec x)$ which is occupied by a large number of bosons and $A_{\vec k} (t)$'s are merely the Fourier mode of this single-particle state. Consequently, the classical field description cannot describe the process of thermalization through which the bosons condensate into the ground state \cite{Sikivie:2009qn, Erken:2011dz, Sikivie:2016enz}.

\subsection{\label{dens_Q}Evolution of density perturbation in quantum field theory}

We assume the potential consists of an external gravitational potential $\hat V_{\rm ext} (\vec x)$ due to a static distribution of baryons with mass density $\rho_B (\vec x)$:
\begin{equation}
\hat V_{\rm ext} (\vec x,t) = (-G) \int_V d^3 x' \ \frac{\rho_B (\vec x')}{|\vec x - \vec x'|}
\end{equation}
and the potential due to the gravitational self-interactions $\hat V_{\rm self}[\psi (\vec x,t)]$ between the axions:
\begin{equation}
\hat V_{\rm self}[\psi (\vec x,t)] = (-Gm) \int_V d^3 x' \ \frac{\psi^\dagger (\vec x',t) \psi (\vec x',t)}{|\vec x - \vec x'|} \ .
\end{equation}
We take $u^{\vec k} (\vec x) $'s to be the momentum eigenstates: $u^{\vec k} (\vec x) = \tfrac{1}{\sqrt{V}} e^{-i\vec k \cdot \vec x}$ and write $a_{\vec k} (t) = e^{-i\omega (\vec k)t} \ b_{\vec k} (t)$ where $\omega(\vec k) = \frac{\vec k \cdot \vec k}{2m}$. Then using Eq.~\ref{eq:SGP_q}, we get the equation of motion for $b_{\vec k} (t)$
\begin{eqnarray}
&&i\partial_t b_{\vec k} = \sum_{\vec k'} e^{it \Omega_{\vec k'}^{\vec k}} \ P(\vec k - \vec k') \ b_{\vec k'} \nonumber \\
&&+ \sum_{\vec k' , \vec k_1, \vec k_2} e^{it \Omega_{\vec k' \vec k_2}^{\vec k \vec k_1} } \ Q(\vec k - \vec k', \vec k_1 - \vec k_2) \ b_{\vec k_1}^\dagger b_{\vec k'} b_{\vec k_2} \label{eq:bkteqn}
\end{eqnarray}
where $\Omega_{\vec k'}^{\vec k} = \omega(\vec k) - \omega(\vec k')$ and $\Omega_{\vec k' \vec k_2}^{\vec k \vec k_1} = \omega(\vec k) + \omega(\vec k_1) - \omega(\vec k') - \omega(\vec k_2)$ and 
\begin{eqnarray}
P(\vec q) &=& \tfrac{(-Gm)}{V} \int_V d^3 x \ d^3 x' \ \frac{e^{i \vec q \cdot \vec x} \rho_B (\vec x')}{|\vec x - \vec x'|} \ \ , \nonumber \\
Q(\vec q_1, \vec q_2) &=& \tfrac{(-Gm^2)}{V^2} \int_V d^3 x \ d^3 x' \ \frac{e^{i \vec q_1 \cdot \vec x + i \vec q_2 \cdot \vec x'}}{|\vec x - \vec x'|}  \ . \label{eq:PQ}
\end{eqnarray}
For a given quantum state $\left| {\rm in} \right>$, the number density is calculated by taking the expectation value of the corresponding operator
\begin{equation}
n_Q (\vec x,t) = \left< {\rm in} \right| \hat n_Q (\vec x,t) \left| {\rm in} \right> = \frac{1}{V} \sum_{\vec k, \vec k'} e^{i (\vec k - \vec k') \cdot \vec x + it\Omega_{\vec k'}^{\vec k}} \  \left< {\rm in} \right| b_{\vec k} ^\dagger (t) b_{\vec k'} (t) \left| {\rm in} \right> \ .
\end{equation}
Since the initial state $\left| {\rm in} \right>$ is associated to $b_{\vec k} ^\dagger (t_0)$'s, one has to solve Eq.~\ref{eq:bkteqn} to find $b_{\vec k} (t)$ in terms of $b_{\vec k} (t_0)$. 

In perturbative expansion, $b_{\vec k} (t)$ can be written as
\begin{equation}
b_{\vec k} (t) = b_{\vec k} ^0 (t) + b_{\vec k} ^1 (t) + b_{\vec k} ^2 (t) + ... 
\end{equation}
where $b_{\vec k} ^n (t)$ contains terms with $n$th power of $G$. Both $P(\vec q)$ and $Q(\vec q)$ are first order terms. Therefore, in the zeroth order of Eq.~\ref{eq:bkteqn}
\begin{equation}
i \partial_t b_{\vec k} ^0 (t) = 0 \ \ \ \implies \ \ \ b_{\vec k} ^0 (t) = b_{\vec k} ^0 (0) = b_{\vec k} (0) \label{eq:zerothbk}
\end{equation}
where we have taken the initial time to be $t_0 = 0$. In the last equality, we have used the fact that the perturbative terms are zero at the initial time i.e. $ b_{\vec k} ^1 (0) = b_{\vec k} ^2 (0) = ... = 0$ (see Eq.~\ref{eq:bk1} for example). Solving Eq.~\ref{eq:bkteqn} in the first order, we get
\begin{widetext}
\begin{eqnarray}
b_{\vec k} ^1 (t) &=& \sum_{\vec k'} (-it) \ {\rm sinc}\big( \tfrac{1}{2}\Omega_{\vec k'}^{\vec k} t \big) e^{\frac{it}{2} \Omega_{\vec k'}^{\vec k}} \ P(\vec k - \vec k') \ b_{\vec k'} (0) \nonumber \\
&+& \sum_{\vec k' , \vec k_1, \vec k_2} (-it) \ {\rm sinc}\big( \tfrac{1}{2} \Omega_{\vec k' \vec k_2}^{\vec k \vec k_1} t \big) e^{\frac{it}{2} \Omega_{\vec k' \vec k_2}^{\vec k \vec k_1}} \ Q(\vec k - \vec k', \vec k_1 - \vec k_2) \ b_{\vec k_1}^\dagger (0) b_{\vec k'} (0) b_{\vec k_2} (0) \ \label{eq:bk1}
\end{eqnarray}
\end{widetext}
where ${\rm sinc} (x) = \frac{\sin x}{x}$. 
The zeroth order term of $\hat n_Q (\vec x,t)$ is
\begin{equation}
\hat n_Q ^0 (\vec x,t) = \frac{1}{V} \sum_{\vec k, \vec k'} e^{i (\vec k - \vec k') \cdot \vec x + it\Omega_{\vec k'}^{\vec k}} \ b_{\vec k} ^{0\dagger} (0) b_{\vec k'} ^0 (0) \label{eq:nQ0}
\end{equation}
where we have used Eq.~\ref{eq:zerothbk}. In the first order, the density operator is given by
\begin{widetext}
\begin{eqnarray}
\hat n_Q ^1 (\vec x,t) &=& \frac{1}{V} \sum_{\vec k, \vec k'} e^{i (\vec k - \vec k') \cdot \vec x + it\Omega_{\vec k'}^{\vec k}} \ \big[b_{\vec k} ^{1\dagger} (t) b_{\vec k'} ^0 (t) + b_{\vec k} ^{0\dagger} (t) b_{\vec k'} ^1 (t) \big] \nonumber \\
&=&  \frac{1}{V} \sum_{\vec k, \vec k' , \vec k''} e^{i (\vec k - \vec k') \cdot \vec x + it\big(\Omega_{\vec k'}^{\vec k} + \frac{1}{2} \Omega_{\vec k''}^{\vec k'} \big)} \ (-it) \ {\rm sinc} \big(\tfrac{1}{2}t \Omega_{\vec k''}^{\vec k'}\big) P(\vec k' - \vec k'') \ b_{\vec k} ^\dagger (0) b_{\vec k''} (0) \nonumber \\
&+& \frac{1}{V} \sum_{\vec k, \vec k' , \vec k'' , \vec k_1 , \vec k_2}  \Big[ e^{i (\vec k - \vec k') \cdot \vec x + it\big(\Omega_{\vec k'}^{\vec k} + \frac{1}{2} \Omega_{\vec k'' \vec k_2}^{\vec k' \vec k_1} \big)} \ (-it) \ {\rm sinc} \big(\tfrac{1}{2}t  \Omega_{\vec k'' \vec k_2}^{\vec k' \vec k_1}\big) \cdot \nonumber \\
&& \ \ \ \  \ \ \ \ \ \ \ \ \cdot Q(\vec k' - \vec k'', \vec k_1 - \vec k_2) \ b_{\vec k} ^\dagger (0) b_{\vec k_1} ^\dagger (0) b_{\vec k''} (0) b_{\vec k_2} (0) \Big] \nonumber \\
&+& {\rm (Hermitian \; conjugates)} \label{eq:nQ1}
\end{eqnarray}
\end{widetext}
where ${\rm sinc} (x) = \frac{\sin x}{x}$. The initial number density $n (\vec x,0)$ is related to the zeroth order by
\begin{equation}
\left< {\rm in} \right| \hat n_Q ^0 (\vec x,0) \left| {\rm in} \right> = n (\vec x,0) 
\end{equation}
where $\left| {\rm in} \right>$ is the intial quantum state of the system. The perturbative terms in the number density are zero at $t=0$ (see Eq.~\ref{eq:nQ1}). In the first order, the contributions from external potential and self-gravitational potential are decoupled, and depend upon the expectation values of $b_{\vec k} ^\dagger (0) b_{\vec k''} (0) $ and $b_{\vec k} ^\dagger (0) b_{\vec k_1} ^\dagger (0) b_{\vec k''} (0) b_{\vec k_2} (0)$, respectively. Up to the linear perturbation, there are no additional terms arising from the commutation of $b_{\vec k}(t)$ and $b_{\vec k}^\dagger (t)$ in quantum field treatment. However, in non-linear regime, such terms will be present and may result into further deviation from classical results.

\subsection{\label{dens_C}Evolution of density perturbation in classical field theory}

In the classical field theory, the wavefunction $\Psi (\vec x,t)$ can be written as
\begin{equation}
\Psi (\vec x,t) = \frac{1}{\sqrt{V}} \sum_{\vec k} e^{-i\vec k \cdot \vec x - i \omega (\vec k)t} \ A_{\vec k} (t)  \label{eq:Psi}
\end{equation}
where $A_{\vec k} (t)$ is a complex number. $A_{\vec k} (t)$ is equivalent to $b_{\vec k} (t)$ in the quantum theory except the fact that $b_{\vec k} (t)$ is an operator. Following the similar steps as in Sec.~\ref{dens_Q}, we calculate the number density. In the zeroth order, the classical number density is
\begin{equation}
n_{\rm cl} ^0 (\vec x,t) = \frac{1}{V} \sum_{\vec k, \vec k'} e^{i (\vec k - \vec k') \cdot \vec x + it\Omega_{\vec k'}^{\vec k}} \ A_{\vec k} ^* (0) A_{\vec k'} (0) \label{eq:ncl0}
\end{equation}
and, in the first order, it is
\begin{widetext}
\begin{eqnarray}
n_{\rm cl} ^1 (\vec x,t) &=& \frac{1}{V} \sum_{\vec k, \vec k' , \vec k''} e^{i (\vec k - \vec k') \cdot \vec x + it\left(\Omega_{\vec k'}^{\vec k} + \frac{1}{2} \Omega_{\vec k''}^{\vec k'} \right)} \ (-it) \ {\rm sinc} \big(\tfrac{1}{2}t \Omega_{\vec k''}^{\vec k'}\big) P(\vec k' - \vec k'') \ A_{\vec k} ^* (0) A_{\vec k''} (0) \nonumber \\
&+& \frac{1}{V} \sum_{\vec k, \vec k' , \vec k'' , \vec k_1 , \vec k_2} \Big[ e^{i (\vec k - \vec k') \cdot \vec x + it\big(\Omega_{\vec k'}^{\vec k} + \tfrac{1}{2} \Omega_{\vec k'' \vec k_2}^{\vec k' \vec k_1} \big)} \ (-it) \ {\rm sinc} \big(\tfrac{1}{2}t  \Omega_{\vec k'' \vec k_2}^{\vec k' \vec k_1}\big) \cdot \nonumber \\
&& \ \ \ \  \ \ \ \ \ \ \ \ \cdot Q(\vec k' - \vec k'', \vec k_1 - \vec k_2) \ A_{\vec k} ^* (0) A_{\vec k_1} ^* (0) A_{\vec k''} (0) A_{\vec k_2} (0) \Big] \nonumber \\
&+& {\rm (complex \; conjugates)} \ . \label{eq:ncl1}
\end{eqnarray}
\end{widetext}
From Eq.~\ref{eq:Psi}, $A_{\vec k} (0)$ can be determined by inverse Fourier transform of the initial wavefunction $\Psi (\vec x,0)$. 

\subsection{\label{dens_Q_C} Why may the classical evolution differ from the quantum evolution?}

We assume that both classical and quantum evolutions start from the same initial values of density and current density. The initial quantum state $\left| {\rm in} \right>$ must satisfy
\begin{eqnarray}
\left< {\rm in} \right| \hat n_Q (\vec x,0) \left| {\rm in} \right> &=& \frac{1}{V} \sum_{\vec k, \vec k'} e^{i(\vec k - \vec k') \cdot \vec x} \left< {\rm in} \right| b_{\vec k}^\dagger (0) b_{\vec k'} (0) \left| {\rm in} \right> = n(\vec x,0) \nonumber \\
\left< {\rm in} \right| \hat {\vec j}_Q (\vec x,0) \left| {\rm in} \right> &=& \frac{(-1)}{2mV} \sum_{\vec k, \vec k'} (\vec k + \vec k') \left< {\rm in} \right| b_{\vec k}^\dagger (0) b_{\vec k'} (0) \left| {\rm in} \right>  = \vec j (\vec x,0) \label{eq:initial_Q} \ .
\end{eqnarray}
Similarly, for the classical field approximation, we must have
\begin{eqnarray}
\frac{1}{V} \sum_{\vec k, \vec k'} e^{i(\vec k - \vec k') \cdot \vec x} A_{\vec k}^* (0) A_{\vec k'} (0) &=& n(\vec x,0) \nonumber \\
\frac{(-1)}{2mV} \sum_{\vec k, \vec k'} (\vec k + \vec k') A_{\vec k}^* (0) A_{\vec k'} (0) &=& \vec j (\vec x,0) \label{eq:initial_C} \ .
\end{eqnarray}
Up to the first order, the classical field treatment yields identical result as the quantum field treatment if 
\begin{equation}
A_{\vec k} ^* (0) A_{\vec k'} (0) = \left< {\rm in} \right| b_{\vec k} ^{\dagger} (0) b_{\vec k'} (0) \left| {\rm in} \right> \label{eq:condition}
\end{equation}
for any $\vec k$, $\vec k'$. In the presence of self-interactions, an additional condition has to be satisfied for classical and quantum evolutions to be identical: 
\begin{equation}
A_{\vec k} ^* (0) A_{\vec k_1} ^* (0) A_{\vec k'} (0) A_{\vec k_2} (0) = \left< {\rm in} \right|  b_{\vec k} ^\dagger (0) b_{\vec k_1} ^\dagger (0) b_{\vec k'} (0) b_{\vec k_2} (0) \left| {\rm in} \right>
\end{equation}
for any $\vec k$, $\vec k'$, $\vec k_1$ and $\vec k_2$. 

We emphasize that Eqs.~\ref{eq:initial_Q}-\ref{eq:initial_C} do not necessarily imply Eq.~\ref{eq:condition}. In the next section we describe a toy model where Eqs.~\ref{eq:initial_Q}-\ref{eq:initial_C} hold but Eq.~\ref{eq:condition} does not hold for a generic quantum state of the system. We explore the differences between the classical evolution and quantum evolutions corresponding to various initial quantum states.

\section{\label{III}Particles in a 1D box}

In this section, we study the quantum and classical evolutions of the number density for a simple and well-defined system. We consider $N$ axions with mass $m$ in a 1D box of length $L$ between $x=0$ and $x=L$. There is a heavy and static point mass $M$ at $x=x_0$. The axions move under the gravitational field of the point mass. We ignore the self-interaction of the axions here. The potential operator is given by
\begin{equation}
\hat V_{\rm ext} (x,t) = \frac{(-GM)}{|x-x_0| + \mu} \label{eq:pot1D}
\end{equation}
where $\mu \ll L$ is introduced as a regulator to avoid the singularities. The axion field is given by
\begin{equation}
\psi (x,t) = \frac{1}{\sqrt{L}} \sum_{k} a_{k} (t) \ e^{-ik x} \label{eq:psi1D}
\end{equation}
where the momenta $k$ are quantized
\begin{equation}
k = \frac{2\pi n_k}{L} \ \ , n_k = 0, \pm1, \pm2, ... \ \ . 
\end{equation}
The expression for $P(q)$ (see Eqs.~\ref{eq:PQ}) is written in terms of dimensionless functions $\tilde P(q)$:
\begin{equation}
P(q) = \tfrac{1}{L} \int_0 ^L dx \ \frac{(-GMm)}{|x - x_0| + \mu} \ e^{iqx} = \left( - \tfrac{GMm}{L} \right) \tilde P (q) \label{eq:tildePQ} 
\end{equation}
The expressions for $\tilde P (q)$ is given in Appendix \ref{A1}.

For the given initial density $n(x,0)$ and velocity field $v (x,0)$, one has to find out the initial wavefunction $\Psi (x,0)$ and $A_{k} (0)$ in the classical field theory. Then the density perturbations can be calculated using Eq.~\ref{eq:ncl1}. In quantum field theory, the initial quantum state must correspond to the initial density $n(x,0)$ and current density $ j (x,0) = n(x,0) v(x,0)$. The density pertubation for that state can be calculated by taking the expectation value of the operators in Eq.~\ref{eq:nQ1}. Here, we consider the scenario where the axion fluid is initially homogeneous with zero current density (hence, zero velocity field) everywhere:
\begin{equation}
n(x,0) = \frac{N}{L} \ \ {\rm and} \ \  j(x,0) = 0 \ . \label{eq:initial1D}
\end{equation}
In Sections \ref{IIIquant} and \ref{IIIclas}, we find evolution of the number density in quantum and classical field treatments, respectively. In Section \ref{IIIcomp}, we compare the results in classical and quantum field theories for certain numerical values of the parameters of the system.

\subsection{\label{IIIquant} Quantum field theory}

We consider the most generic quantum state (see Eq.~\ref{eq:nQt0}) corresponding to homogeneous number density and zero velocity field:
\begin{equation}
\left| {\rm in} \right> = \left| N_1 (p_1), N_2 (p_2), N_3 (p_3), ... \right>  \label{eq:states}
\end{equation}
where there are $N_i$ particles in momentum state $k=p_i$ ($i = 1, 2, 3, ....$) and 
\begin{equation}
\sum_i N_i = N \ \ {\rm and} \ \ \sum_i N_i p_i = 0 \ .
\end{equation}
The last equality ensures $\left< {\rm in} \right| \hat j_Q (x, 0) \left| {\rm in} \right> = 0$. In the zeroth order, the number density remains constant:
\begin{equation}
n_Q^{0} (x,t) = n(x,0) = \frac{N}{L} \ .
\end{equation}
In the first order, the density contrast, $\delta (x,t) = \frac{1}{n(x,0)}[\langle \hat n^1 (x,t) \rangle - n(x,0)]$ corresponding to the state $\left| {\rm in} \right>$ becomes
\begin{equation}
\delta_Q (x,t) = 2 \Big( \frac{GMmt}{L} \Big) \sum_i f_i \ \tilde F (p_i , x, t)  \label{eq:nQ1D}
\end{equation}
where $f_i = \frac{N_i}{N}$ and $\tilde F (q , x, t)$ is a dimensionless function given by
\begin{widetext}
\begin{equation}
\tilde F (q , x, t) = \sum_k {\rm sinc} \big(\frac{1}{2} \Omega_q^k t\big) \Big[ - \cos \big( (k-q)x + \frac{1}{2} \Omega_q^k t \big) \tilde P_I (k-q) 
+ \sin \big( (k-q)x + \frac{1}{2} \Omega_q^k t \big) \tilde P_R (k-q) \Big] \label{eq:Fext}
\end{equation}
\end{widetext}
with ${\rm sinc} (x) = \tfrac{\sin x}{x}$. The expressions for $\tilde P (q)$'s are given by Eqs.~\ref{eq:PRtilde}-\ref{eq:PItilde}. As shown by Eq.~\ref{eq:nQ1D}, each single-particle state that is occupied, contributes to the density perturbation.

\subsection{\label{IIIclas} Classical field theory}

For the initial conditions in Eqs.~\ref{eq:initial1D}, the classical wavefunction $\Psi (x,0)$ is determined upto a constant phase:
\begin{equation} 
\Psi(x,0) = \sqrt{\frac{N}{L}} \implies A_k (0) = \sqrt{N} \delta_{k,0} \ .
\end{equation}
The number density in the zeroth order is
\begin{equation}
n_{\rm cl}^{0} (x,t) = \frac{N}{L}  \ .
\end{equation}
The density contrast in the first order is found to be
\begin{equation}
\delta_{\rm cl} (x,t) = 2 \frac{GMmt}{L} \  \tilde F (p=0 , x, t) \ . \label{eq:ncl1D}
\end{equation}
The above equation is equivalent to the quantum field result, Eq.~\ref{eq:nQ1D}, with all the particles in $p_1 = 0$ state ($f_1 = 1$ and all other $f_i$'s are zero).

\subsection{\label{IIIcomp} Quantum vs classical evolution}

For initially homogeneous system of axions with zero velocity field, the generic quantum evolution of the density contrast is given by Eq.~\ref{eq:nQ1D} and the corresponding classical evolution is given by Eq.~\ref{eq:ncl1D}.

(1) When $\Omega_q^k t \ll 1$ in Eq.~\ref{eq:Fext}, ${\rm sinc} \big(\frac{1}{2} \Omega_q^k t\big) \approx 1$ and, $\tilde F(q,x,t)$ is independent of $q$. Then Eqs.~\ref{eq:nQ1D} and \ref{eq:ncl1D} are almost equal, i.e. the quantum  and classical evolutions closely follow each other. In this regime, the evolution does not depend upon the initial quantum state of the system. 

(2) When $\Omega_q^k t \gg 1$, ${\rm sinc} \big(\frac{1}{2} \Omega_q^k t\big) \sim \delta(\Omega_q^k t)$ and, $\tilde F (q,x,t)$ depends upon $q$. Therefore, in this regime, different initial quantum states corresponding to the same initial density and velocity field, may result into different number density and velocity field.

The quantum evolution depends upon the initial quantum state of the system and, in general, differ from the unique classical evolution after a certain time $t = t_{\rm classical}$ defined by
\begin{equation}
\Omega_k^p t_{\rm classical} = (n_p ^2 - n_k ^2) \frac{2 \pi ^2 t_{\rm classical}}{mL^2} = 0.1 (n_p ^2 - n_k ^2)  \label{eq:tclass}
\end{equation}
where we have used $p = \frac{2\pi n_p}{L}$ and $k = \frac{2\pi n_k}{L}$. The fraction $0.1$ is somewhat arbitrary. The required condition is that the typical value of $\Omega_k^p t$ is of $\mathcal O (1)$. Here, we express the momentum $p$ in terms of the state $n_p$ of the particle. So $p$ is independent of the mass $m$. Consequently, the energy difference $\Omega_k^p$ is inversely proportional to $m$ when we change the mass $m$ keeping the particle state $n_p$ fixed. If we were to keep the velocity $v_{n_p} = 2\pi n_p /(mL) $ fixed, the energy difference $\Omega_k^p$ should have been proportional to the mass $m$.

\subsection{\label{IIInum} Examples with numerical values}

	For a 1D box of length $L$, the momenta of the particles are given by: $k_n = 2\pi n/L$ and velocities are $v_n = 2\pi n/(mL)$. Our calculations are valid for non-relativistic bosons, therefore we must choose the numerical values such that $v_n \ll 1$ {\it i.e.} $mL \gg 1$. For $m = 10^{-20}$ eV and length of the box as $L = 1$ pc, the velocities are of the order of 
\begin{equation}
\frac{1}{mL} = 6.2 \times 10^{-4} \Big(\frac{10^{-20} \ {\rm eV}}{m}\Big) \Big(\frac{1 \ {\rm pc}}{L}\Big)
\end{equation}
which are well within the non-relativistic regime. For a time-scale of 1 Myr, we get
\begin{equation}
mt = 4.8 \times 10^{8} \Big(\frac{m}{10^{-20} \ {\rm eV}}\Big) \Big( \frac{t}{1 \ {\rm Myr}} \Big) \ .
\end{equation}
Following Eq.~\ref{eq:tclass}, the timescale of the validity of classical field approximation is
\begin{equation}
t_{\rm classical} = 2.7 \times 10^{-5} \ {\rm Myr} \Big(\frac{m}{10^{-20} \ {\rm eV}}\Big) \Big( \frac{L}{1 \ {\rm pc}} \Big)^2 \ . \label{eq:tclassnum}
\end{equation}
The above equation holds for all values of $m$ and $L$ as well as for any external potential if the system is in the regime of linear perturbation.

\begin{figure}
\includegraphics[scale=0.3]{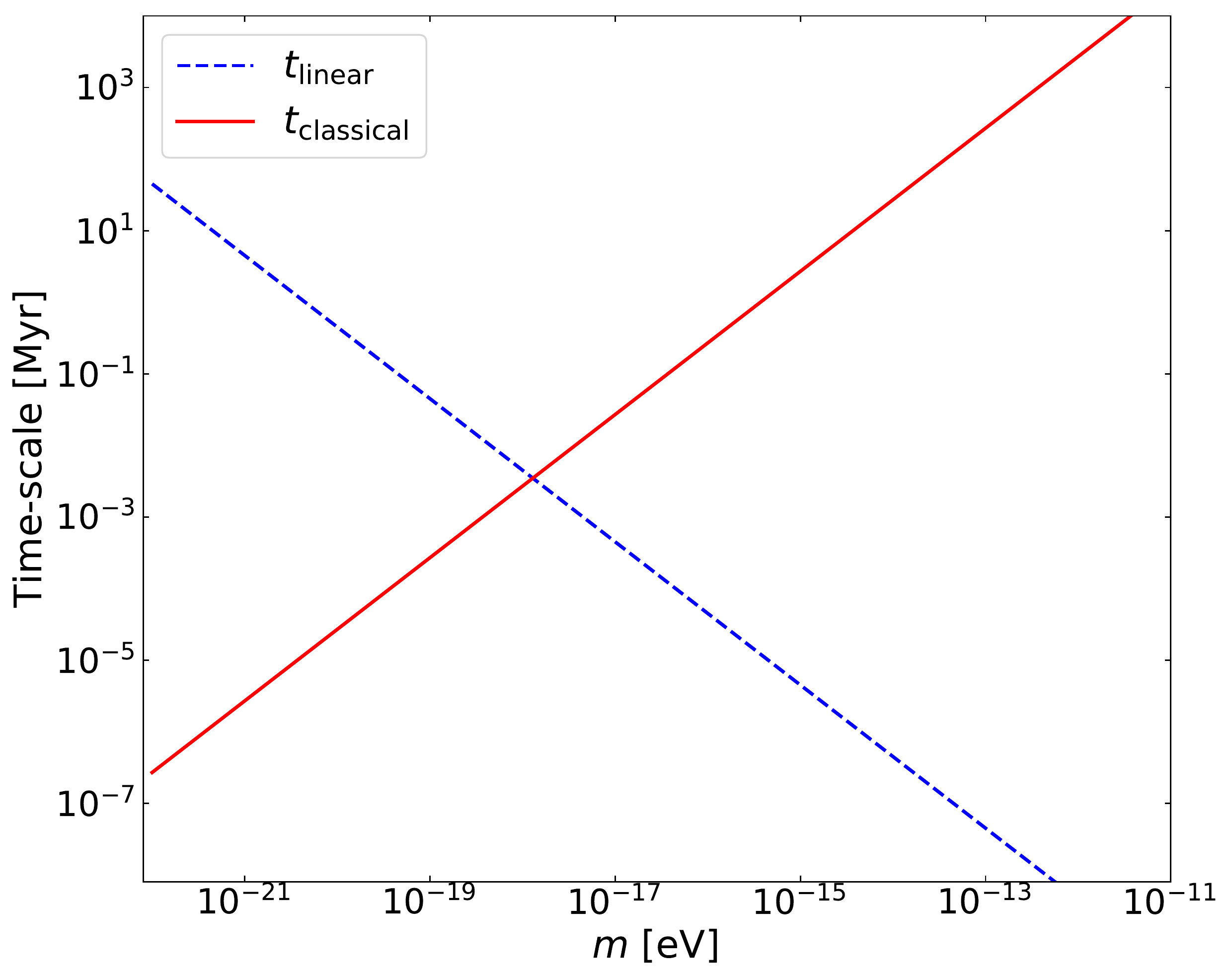}
\caption{\label{fig:timescale} Time-scale of the validity of linear perturbation, $t_{\rm linear}$ (see Eq.~\ref{eq:tlinnum}), and the time-scale of the validity of classical field approximation, $t_{\rm classical}$ (see Eq.~\ref{eq:tclassnum}) as functions of $m$. We have taken for $M = M_\odot$ and $L = 1$ pc.}
\end{figure}

The magnitude of density perturbation is governed by $(GMmt/L)$. We define the timescale of validity of the linear perturbation $t = t_{\rm linear}$ such that 
\begin{equation}
\frac{GMmt_{\rm linear}}{L} = 10^{-5} \ .
\end{equation}
We choose the static point mass to be $M = M_\odot$ which yields
\begin{equation}
t_{\rm linear} = 0.45 \ {\rm Myr} \Big(\frac{10^{-20} \ {\rm eV}}{m}\Big) \Big( \frac{M_\odot}{M} \Big) \Big( \frac{L}{1 \ {\rm pc}} \Big) \ . \label{eq:tlinnum}
\end{equation}
We place the point mass $M$ at $x_0 = 0.6 L$ and take the regulator as $\mu = 10^{-13} L$. Figure \ref{fig:timescale} shows $t_{\rm linear}$ and $t_{\rm classical}$ as a function of $m$ for $M = M_\odot$, $L = 1$ pc. For $m \lesssim 10^{-18}$ eV, $t_{\rm classical}$ is less than $t_{\rm linear}$, i.e. the quantum evolution of the number density differs from the classical evolution within the linear regime.

Here we show how different initial quantum states corresponding to the same density and velocity field, may result into different evolutions of the density contrast. We consider the following initial quantum states written in the form $\vert f_1 (n_1), f_2 (n_2), ... \rangle$ where $f_i = \frac{N_i}{N}$ and $p_i = \frac{2\pi n_i}{L}$:
\begin{eqnarray}
\vert 1 \rangle &=& \vert 1.0 (0), 0, 0, ... \rangle \nonumber \\
\vert 2 \rangle &=& \vert 0.9 (0), 0.05 (-2), 0.05 (2), 0, 0, ... \rangle \nonumber \\ 
\vert 3 \rangle &=& \vert 0.5 (0), 0.3 (2), 0.15 (-3), 0.15 (-1), 0, 0, ... \rangle . \label{eq:Qstates}
\end{eqnarray}
Since $\sum f_i = 1$ and $\sum f_i n_i = 0$, all of these quantum states correspond to the same initial condition i.e., Eqs.~\ref{eq:initial1D}. We emphasize that, in classical field theory, the initial wavefunction is uniquely specified by the initial conditions, Eqs.~\ref{eq:initial1D}. If we write the classical wavefunction corresponding to the quantum states $\vert 2 \rangle$ and $\vert 2 \rangle$, we get inhomogeneous density and non-zero velocity field which result into entirely different evolutions of the number densitites (see the discussion in Sec.~\ref{inconsistency}). 

\begin{figure*}
\includegraphics[scale=0.3]{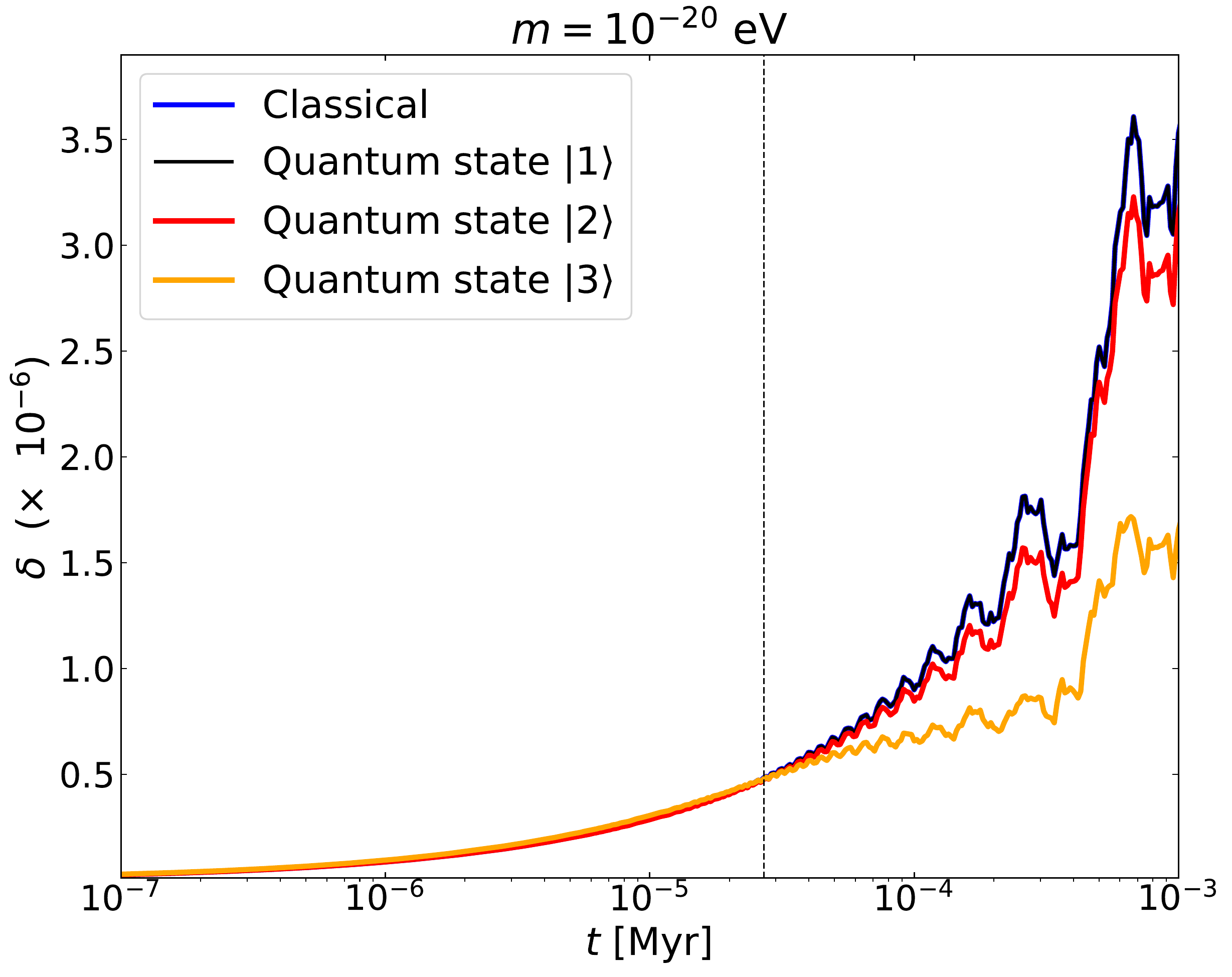}
\includegraphics[scale=0.3]{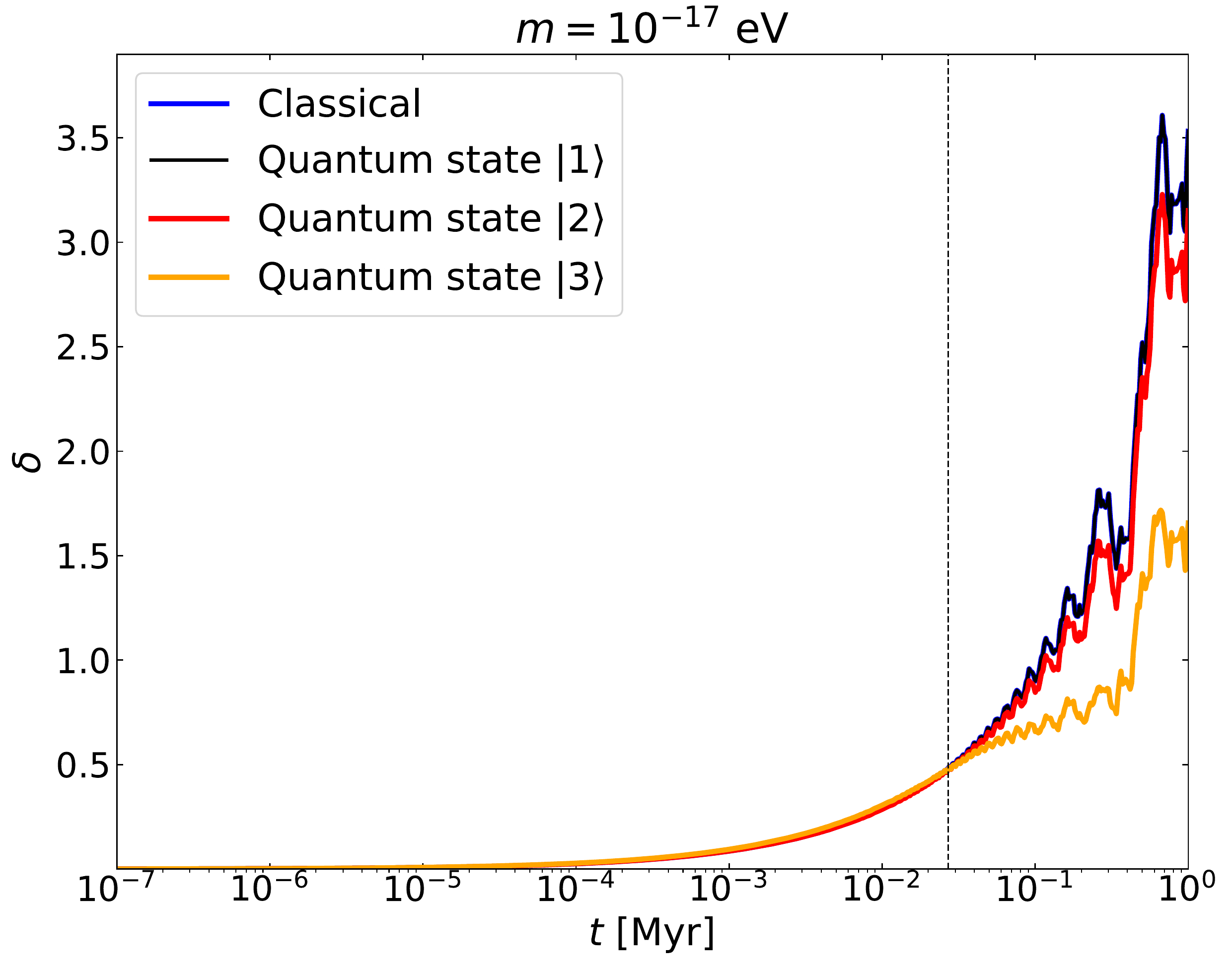}
\caption{\label{fig:deltavstau1} {\it Left panel:} Variation of the density contrast $\delta$ at a specific point $x = 0.6 L$ in classical field theory (see Eq.~\ref{eq:ncl1D}) and in quantum field theory for different initial quantum states (see Eq.~\ref{eq:nQ1D}). Quantum states $\vert 1 \rangle$, $\vert 2 \rangle$ and $\vert 3 \rangle$ are defined in Eq.~\ref{eq:Qstates}. Quantum evolution corresponding to the state $\vert 1 \rangle$ is identical to the classical evolution. In this example, we have chosen $m = 10^{-20}$ eV, $M = M_\odot$ and $L = 1$ pc. The vertical dashed line indicates the timescale for the validity of classical field approximation, $t_{\rm classical} = 2.7 \times 10^{-5}$ Myr (see Eq.~\ref{eq:tclassnum}). {\it Right panel:} The same as the left panel except for $m = 10^{-17}$ eV. The vertical dashed line corresponds to $t_{\rm classical} = 2.7 \times 10^{-2}$ Myr (see Eq.~\ref{eq:tclassnum}). }
\end{figure*}

Quantum evolution of the number density corresponding to the state $\vert 1 \rangle$ is the same as the classical evolution. In the left panel of Fig.~\ref{fig:deltavstau1}, we show the evolution of density contrast $\delta$ at a specific point $x= 0.6 L$ in classical  (Eq.~\ref{eq:ncl1D}) and quantum field theories (Eq.~\ref{eq:nQ1D}), for $m = 10^{-20}$ eV. As expected from Eq.~\ref{eq:tclassnum}, the classical and quantum evolutions differ from each other at about $t = 2.7 \times 10^{-5}$ Myr. In the right panel of Fig.~\ref{fig:deltavstau1}, we show the same for $m = 10^{-17}$ eV, and the evolutions differ from each other at about $t = 2.7 \times 10^{-2}$ Myr as predicted by Eq.~\ref{eq:tclassnum}. However, at that time, the density evolutions enter non-linear regime ($t_{\rm linear} = 4.5 \times 10^{-4}$ Myr for $m = 10^{-17}$ eV following Eq.~\ref{eq:tlinnum}) where our analytical expressions are not valid.

\section{\label{IV}Summary}

In classical field theories, the axions are represented by a complex wavefunction which satisfies the Schro\"dinger-Gross-Pitaevskii equation in the non-relativistic regime. The number density and velocity field of axions follow the continuity equation and Euler-like equation with an additional `quantum pressure' term. The classical evolution is uniquely determined once the initial density and velocity field are specified. We write down the analogues of the continuity and Euler-like equations in quantum field theory. We show that, in quantum field treatment, the evolution of axions is uniquely specified only if the initial quantum state of the system is known. There exist a large number of quantum states corresponding to the same density and velocity field. Evolutions of number density for these quantum states are not identical and, in general, differ from the classical evolution.

We consider a toy-model with a large number of axions in a one-dimensional box and moving under the gravitational potential of a static point mass. Initially, the number density is homogeneous and the velocity field is zero everywhere. Here we ignore the self-interactions between the axions. We show that, after a certain time, the quantum evolution of the number density depends upon the initial quantum state of the system and differs from the classical evolution. We give an estimate of the timescale of the validity of classical field approximation. We also show that, for certain parameters of the system, the classical and quantum evolutions of density perturbation may differ from each other within the linear regime. We note that the differences between classical and quantum evolutions depend upon what fraction of quanta is in a certain single-particle state, but not on the absolute number of axions.

In a future work, we will discuss whether the difference between classical and quantum field evolutions of density perturbation is significant for cosmological axions. For such studies, it is important to include the expansion of the Universe as well as the effects of self-interactions between the axions.


\begin{acknowledgments}

I gratefully acknowledge the stimulating discussions with Pierre Sikivie. I am also thankful to Elisa Todarello, Ariel Arza, Raghavan Rangarajan, Koushik Datta and Basudeb Dasgupta for many engaging discussions. SSC is supported by the grant ``The Milky Way and Dwarf Weights with Space Scales'' funded by University of Torino and Compagnia di S. Paolo (UniTO-CSP), by the grant no. IDROL 70541 IDRF 2020.0756 funded by Fondazione CRT, and by INFN. We acknowledge partial support from the INFN grant InDark and the Italian Ministry of Education, University and Research (MIUR) under the Departments of Excellence grant L.232/2016.

\end{acknowledgments}

\bibliography{axion_bib.bib}

\clearpage

\appendix

\section{\label{A1}Expressions for $\tilde P (q)$}

$\tilde P(q)$ is defined in Eqs.~\ref{eq:tildePQ} as:
\begin{equation}
\tilde P(q) = \int_0 ^L dx \ \frac{1}{|x - x_0| + \mu} \ e^{iqx}  \ .
\end{equation}
The expressions for the real and imaginary parts of $\tilde P (q) = \tilde P_R (q) + \tilde P_I (q) $ are
\begin{widetext}
\begin{eqnarray}
\tilde P_R (q\neq0) &=&  \cos(qx_0) \Big[{\rm Ci}(qx_0 + q\mu) + {\rm Ci} (qL - qx_0 + q\mu) - 2{\rm Ci}(q\mu) \Big] \nonumber \\
&&- \sin(qx_0) \Big[ {\rm Si} (qL - qx_0) - {\rm Si}(qx_0) \Big] \nonumber \\
\tilde P_R (q=0) &=& 2 \ln \Big( \frac{\sqrt{x_0 (1-x_0)}}{\mu} \Big) \label{eq:PRtilde}
\end{eqnarray}
and 
\begin{eqnarray}
\tilde P_I (q\neq0) &=&  \sin(qx_0) \Big[{\rm Ci}(qx_0 + q\mu) + {\rm Ci} (qL - qx_0 + q\mu) - 2{\rm Ci}(q\mu) \Big] \nonumber \\
&&- \cos(qx_0) \Big[ {\rm Si} (qL - qx_0) - {\rm Si}(qx_0) \Big] \nonumber \\
\tilde P_I (q=0) &=& 0 \ . \label{eq:PItilde}
\end{eqnarray}
\end{widetext}
${\rm Ci} (x)$ and ${\rm Si} (x)$ are integrals of $\frac{\cos x}{x}$ and $\frac{\sin x}{x}$ respectively:
\begin{eqnarray}
{\rm Ci} (b) - {\rm Ci} (a) &=& \int_a ^b dx \frac{\cos x}{x} \nonumber \\
{\rm Si} (b) - {\rm Si} (a) &=& \int_a ^b dx \frac{\sin x}{x}  \ .
\end{eqnarray}

\clearpage

\end{document}